\documentclass[12pt]{article}
\usepackage{latexsym,amssymb,amsfonts,amsmath}
\usepackage{mathrsfs}
\usepackage[dvips]{graphicx}
\newlength{\extraspace}
\newlength{\extraspaces}
\newcommand{\be}{\begin{equation}
\addtolength{\abovedisplayskip}{\extraspaces}
\addtolength{\belowdisplayskip}{\extraspaces}
\addtolength{\abovedisplayshortskip}{\extraspace}
\addtolength{\belowdisplayshortskip}{\extraspace}}
\newcommand{\ee}{\end{equation}}

\newcommand{\bea}{\begin{eqnarray}
\addtolength{\abovedisplayskip}{\extraspaces}
\addtolength{\belowdisplayskip}{\extraspaces}
\addtolength{\abovedisplayshortskip}{\extraspace}
\addtolength{\belowdisplayshortskip}{\extraspace}}
\newcommand{\eea}{\end{eqnarray}}

\newcommand{\newsection}[1]{
\pagebreak[3]
\addtocounter{section}{1}
\setcounter{equation}{0}
\setcounter{subsection}{0}
\setcounter{footnote}{0}
\begin{flushleft}
{\large\bf \thesection. #1}
\end{flushleft}
\nopagebreak
\nopagebreak}

\newcommand{\La}{\mathcal{L}}

\newcommand{\tc}{T_{c}}
\newcommand{\tuone}{T_{U(1)}}

\newcommand{\tro}{T_{\rho_{\pi}}}
\newcommand{\ov}[1]{\overline{#1}}
\newcommand{\gr}[1]{\textbf{#1}}

\newcommand{\lsigma}{\mathcal{L}_{0}(U,U^{\dagger})}
\newcommand{\ropi}{\rho_{\pi}}
\newcommand{\rox}{\rho_{X}\,}
\newcommand{\bra}{\langle}
\newcommand{\ket}{\rangle}

\newcommand{\lpq}{\lambda_{\pi}^{2}}
\newcommand{\Lpq}{\Lambda_{\pi}^{2}}
\newcommand{\apq}{A_{\pi}^{2}}
\newcommand{\bpq}{B_{\pi}^{2}}
\newcommand{\lxq}{\lambda_{X}^{2}}
\newcommand{\fxq}{F_{X}^{2}}
\newcommand{\rad}{\sqrt}
\newcommand{\de}{\partial}
\newcommand{\qq}{\langle\ov{q}_l q_l\rangle}
\newcommand{\fpq}{F_{\pi}^{2}}
\newcommand{\Gpi}{\mathcal{G}_{\pi}}

\newcommand{\smin}{\ov{\sigma}}
\newcommand{\dmin}{\ov{\delta}}
\newcommand{\amin}{\ov{\alpha}}

\newcommand{\Ord}{\mathcal{O}}
\newcommand{\Tr}{\mathrm{Tr}}

\begin{document}

\addtolength{\baselineskip}{.8mm}

{\thispagestyle{empty}

\noindent \hspace{1cm} \hfill IFUP--TH/2013--17 \hspace{1cm}\\

\begin{center}
{\large\bf REMARKS ON THE $U(1)$ AXIAL SYMMETRY}\\
{\large\bf AND THE CHIRAL TRANSITION IN QCD}\\
{\large\bf AT FINITE TEMPERATURE}\\
\vspace*{1.0cm}
{\large
Enrico Meggiolaro$^{1,}$\footnote{E-mail: enrico.meggiolaro@df.unipi.it}
and
Alessandro Mord\`a$^{2,3,}$\footnote{E-mail: morda@cpt.univ-mrs.fr}
}\\
\vspace*{0.5cm}{\normalsize
{$^{1}$ Dipartimento di Fisica, Universit\`a di Pisa,
and INFN, Sezione di Pisa,\\
Largo Pontecorvo 3, I-56127 Pisa, Italy}}\\
\vspace*{0.5cm}{\normalsize
{$^{2}$ CPPM, Aix--Marseille Universit\'e, CNRS/IN2P3,
F-13288 Marseille, France}}\\
\vspace*{0.5cm}{\normalsize
{$^{3}$ CPT, Aix--Marseille Universit\'e and Universit\'e du Sud Toulon--Var,\\
CNRS (UMR 7332), F-13288 Marseille, France}}\\
\vspace*{2cm}{\large \bf Abstract}
\end{center}

\noindent
We discuss the role of the $U(1)$ axial symmetry for the scalar and
pseudoscalar meson mass spectrum of QCD at finite temperature, above
the chiral transition at $\tc$, using a chiral effective Lagrangian model,
which, in addition to the usual chiral condensate
$\langle \bar{q} q \rangle$, also includes a (possible)
{\it genuine} $U(1)_A$-breaking condensate that (possibly) survives across
the chiral transition.
The motivations for considering this Lagrangian (and a critical comparison
with other effective Lagrangian models existing in the literature) are
presented.
A detailed comparison between the case $N_f\geq 3$ and the (remarkably
different) case $N_f=2$ is performed. The results obtained in the case
$N_f=2$ are also critically compared with the available lattice results.

\vspace{0.5cm}
}
\newpage

\newsection{Introduction}

\noindent
It is well known that, at zero temperature, the $SU(N_f) \otimes SU(N_f)$
chiral symmetry of the QCD Lagrangian with $N_f$ massless quarks
(the physically relevant cases being $N_f=2$ and $N_f=3$)
is spontaneously broken down to the vectorial subgroup $SU(N_f)_V$ by the
condensation of $q\bar{q}$ pairs, i.e., by the nonzero value of the
vacuum expectation value $\langle \bar{q}q \rangle
\equiv \sum_{l=1}^{N_f} \langle \bar{q}_l q_l \rangle$ (the so-called
{\it chiral condensate}), and the $N_f^2-1$ $J^P=0^-$ mesons
are just the Goldstone bosons associated with this breaking
(see, e.g., Ref. \cite{chiral-symmetry} and references therein).
One expects that this scenario not only holds for massless quarks, but also
continues for a small quark mass region, in which the Goldstone bosons become
{\it pseudo-Goldstone} bosons, with small (if compared with other
hadrons) nonzero masses.
The chiral condensate $\langle \bar{q}q \rangle$ is an order parameter
for the chiral symmetry breaking: at high temperatures, the thermal energy
breaks up the $q\bar{q}$ condensate, leading to the restoration of chiral
symmetry for temperatures above the chiral phase transition temperature
$T_{c}$, defined as the temperature at which the chiral condensate
$\langle \bar{q}q \rangle$ goes to zero (in the {\it chiral limit}
$m_1 = \dots = m_{N_f} = 0$).
From lattice determinations of $\langle \bar{q}q \rangle$, it is known (see,
e.g., Refs. \cite{HotQCD}) that this critical temperature is of the order
$\tc \sim 150 \div 170$ MeV and practically equal to the {\it deconfinement}
temperature $T_{d}$, separating the {\it confined} (or {\it hadronic})
phase at $T<T_d$, from the {\it deconfined} phase (also known as
{\it quark-gluon plasma}) at $T>T_d$.
But this is not the whole story, since, in addition to the
$SU(N_f) \otimes SU(N_f)$ chiral symmetry, QCD with $N_f$ massless quarks
also has a $U(1)$ axial symmetry (at least at the classical level)
\cite{Weinberg1975,tHooft1976}. This symmetry is broken by an anomaly at the
quantum level, which in the Witten--Veneziano mechanism
\cite{Witten1979,Veneziano1979} plays a fundamental role (via the so-called
{\it topological susceptibility}) in explaining the large mass of the
$\eta'$ meson.

The role of the $U(1)$ axial symmetry for the finite temperature phase 
structure has been not well understood so far. One expects that at very high
temperatures also the $U(1)$ axial symmetry will be ({\it effectively})
restored (since, at least for $T \gg T_c$, the density [in the partition
function] of the instanton configurations, responsible for the $U(1)_A$
breaking, are strongly suppressed due to a Debye-type screening
\cite{GPY1981}); but it is still an open question of hadronic physics whether
the fate of the $U(1)$ chiral symmetry of QCD has or has not something to do
with the fate of the $SU(N_f) \otimes SU(N_f)$ chiral symmetry.
This question is surely of phenomenological relevance since the particle
mass spectrum above $T_c$ drastically depends on the presence or absence of the
$U(1)$ axial symmetry. From the theoretical point of view, this question
can be investigated by comparing the behavior at nonzero temperatures of the
two-point correlation functions for the following $q\bar{q}$ meson channels
(we consider for simplicity the case of $N_f=2$ light flavors)
\cite{DK1987,Shuryak1994}:
the isoscalar ($I=0$) scalar channel $\sigma$ (also known as $f_0$ in the
modern language of hadron spectroscopy), interpolated by the operator
$O_\sigma = \bar{q} q$;
the isovector ($I=1$) scalar channel $\vec\delta$
(also known as $\vec{a}_0$), interpolated by the operator
$\vec{O}_\delta = \bar{q} \frac{\vec{\tau}}{2} q$;
the isoscalar ($I=0$) pseudoscalar channel $\eta$, interpolated by the operator
$O_{\eta} = i\bar{q} \gamma_5 q$;
and the isovector ($I=1$) pseudoscalar channel $\vec\pi$, interpolated by the
operator $\vec{O}_\pi = i\bar{q} \gamma_5 \frac{\vec{\tau}}{2} q$.
Under $SU(2)_A$ transformations, $\sigma$ is mixed with $\vec\pi$r;: thus, the
restoration of this symmetry at $T_{c}$ requires identical correlators
for these two channels, which implies, in particular, identical
{\it chiral susceptibilities}, $\chi_\sigma = \chi_\pi$
[$\chi_f \equiv \int d^4x~ \langle T O_f(x) O_f^\dagger(0) \rangle$],
and identical ({\it screening}) masses, $M_\sigma = M_\pi$.
Another $SU(2)$ chiral multiplet is $(\eta,\vec\delta)$.
On the contrary, under $U(1)_A$ transformations, $\vec\delta$ is mixed
with $\vec\pi$, so, an {\it effective restoration} of the $U(1)$ axial
symmetry should imply that these two channels become degenerate, with
identical correlators and, therefore, with identical chiral susceptibilities,
$\chi_\delta = \chi_\pi$, and identical (screening) masses, $M_\delta = M_\pi$.
Another $U(1)$ chiral multiplet is $(\sigma,\eta)$.
(Clearly, if both chiral symmetries are restored, then all $\sigma$, $\vec\pi$,
$\eta$, and $\vec\delta$ correlators should become the same.)

In this paper, we shall analyze the scalar and pseudoscalar meson
mass spectrum, above the chiral transition at $\tc$, using a chiral
effective Lagrangian model (which was originally proposed in
Refs. \cite{EM1994a,EM1994b,EM1994c} and elaborated on in Refs.
\cite{MM2003,EM2004,EM2011}), which, in addition to the usual
chiral condensate $\langle \bar{q} q \rangle$, also includes a (possible)
{\it genuine} $U(1)_A$-breaking condensate that (possibly) survives across
the chiral transition at $T_{c}$, staying different from zero at $T > T_{c}$.
The motivations for considering this Lagrangian (and a critical comparison
with other effective Lagrangian models existing in the literature) are
presented in Sec. 2.
The results for the mesonic mass spectrum for $T > T_{c}$ are derived in
Sec. 3, for the case $N_f\geq 3$, and in Sec. 4, for the case $N_f=2$.
Finally, in Sec. 5, we shall summarize the results that we have obtained
and we shall make some comments on (i) the remarkable difference between
the case $N_f\geq 3$ and the case $N_f=2$ and (ii) the comparison
between our results and the available lattice results for $N_f=2$.

\newsection{Chiral effective Lagrangians}

\noindent
Chiral symmetry restoration at nonzero temperature is often studied in the
framework of the following effective Lagrangian
\cite{PW1984,LRS2000,RRR2003,Vicari-et-al.}
(which had been originally proposed to study the chiral dynamics
at $T=0$ \cite{Levy1967,BL1969,GG1969}),
written in terms of the (quark-bilinear) mesonic effective field
$U_{ij} \sim \ov{q}_{jR} q_{iL} =
\ov{q}_{j}\left(\frac{1+\gamma^{5}}{2}\right)q_{i}$
(up to a multiplicative constant),\footnote{Throughout this paper, we use
the following notations for the left-handed and right-handed quark fields:
$q_{L,R} \equiv \frac{1}{2} (1 \pm \gamma_5) q$,
with $\gamma_5 \equiv -i\gamma^0\gamma^1\gamma^2\gamma^3$.}
\begin{equation}\label{lageffstandard}
\mathcal{L}_1(U,U^\dag)=\mathcal{L}_0(U,U^{\dagger})+
\frac{B_{m}}{2\sqrt2}\Tr[MU+M^{\dagger}U^{\dagger}]+\mathcal{L}_{I}(U,U^\dag),
\end{equation}
where $\mathcal{L}_{0}(U,U^{\dagger})$ describes a kind of linear
sigma model:
\begin{eqnarray}\label{sigmamodel}
\mathcal{L}_0(U,U^{\dagger}) &=&
\frac{1}{2}\Tr[\partial_{\mu}U\partial^{\mu}U^{\dagger}]- V_0(U,U^\dagger),
\nonumber \\
V_0(U,U^\dagger) &=&
\frac{1}{4}\lambda_{\pi}^{2}\Tr[(UU^{\dagger}-\rho_{\pi}\mathbf{I})^{2}]
+\frac{1}{4}\lambda_{\pi}^{\prime 2}[\Tr(UU^{\dagger})]^{2}.
\end{eqnarray}
{\bf I} is the identity matrix, $M={\rm diag}(m_1,\dots,m_{N_f})$
represents the quark mass matrix, which enters in the QCD Lagrangian as
$\delta {\cal L}^{(mass)}_{QCD} = -\bar{q}_R M q_L -\bar{q}_L M^\dagger q_R$,
while $\mathcal{L}_{I}(U,U^\dag)$ is an interaction term of the form:
\begin{equation}\label{thooftterm}
\mathcal{L}_{I}(U,U^\dagger)=c_{I}[\det U+\det U^{\dagger}].
\end{equation}
Since under $U(N_f)_L\otimes U(N_f)_R$ chiral transformations the quark fields
and the mesonic effective field $U$ transform as
\begin{equation}\label{trasfU}
U(N_f)_L\otimes U(N_f)_R:~~~~
q_{L}\rightarrow V_{L}q_{L}, ~~q_{R}\rightarrow V_{R}q_{R}~~\Rightarrow~~
U\rightarrow V_{L} U V_{R}^{\dag},
\end{equation}
where $V_L$ and $V_R$ are arbitrary $N_f \times N_f$ unitary matrices,
we have that $\lsigma$ is invariant under the entire chiral group
$U(N_f)_L\otimes U(N_f)_R$, while the interaction term \eqref{thooftterm} [and
so the entire effective Lagrangian \eqref{lageffstandard} in the chiral limit
$M=0$] is invariant under $SU(N_f)_L\otimes SU(N_f)_R \otimes U(1)_V$ but
{\it not} under a $U(1)$ axial transformation:\footnote{For the case of
$N_f=2$ flavors, two other four-point couplings with the same property,
i.e., invariant under $SU(N_f)_L\otimes SU(N_f)_R \otimes U(1)_V$ but not
under $U(1)_A$, could be considered \cite{PW1984,Vicari-et-al.}; however,
these terms are not relevant for the type of analysis that we are going
to perform in this paper.}
\begin{equation}\label{U1trasf}
U(1)_A:~~~~q_L\rightarrow e^{-i\alpha}q_L, ~~q_R\rightarrow e^{i\alpha}q_R
~~\Rightarrow~~ U \rightarrow e^{-i2\alpha} U .
\end{equation}
It is often claimed (see, for example, Ref. \cite{tHooft1986} and references
therein) that {\it instanton} processes, which are known to break the $U(1)_A$
symmetry by means of an effective $2N_f$-quark vertex that is invariant
under $SU(N_f)_L\otimes SU(N_f)_R \otimes U(1)_V$, but not under a $U(1)$ axial
transformation, can be modelled using the interaction term \eqref{thooftterm}.

However, as was noticed by Witten \cite{Witten1980}, Di Vecchia, and Veneziano
\cite{DV1980}, this type of {\it anomalous term} does not correctly
reproduce the U(1) axial anomaly of the fundamental theory, i.e., of the QCD
(and, moreover, it is inconsistent with the $1/N_c$ expansion).
In fact, one should require that, under a $U(1)$ axial transformation
\eqref{U1trasf}, the effective Lagrangian, in the chiral limit $M=0$,
transforms as
\begin{equation}\label{trasfanomal}
U(1)_A:~~~~ \mathcal{L}^{(M=0)}_{eff}(U,U^\dagger,Q)\rightarrow
\mathcal{L}^{(M=0)}_{eff}(U,U^\dagger,Q) + \alpha 2N_f Q ,
\end{equation}
where $Q(x)=\frac{g^{2}}{64\pi^{2}}\varepsilon^{\mu\nu\rho\sigma}
F_{\mu\nu}^{a}(x)F_{\rho\sigma}^{a}(x)$ is the {\it topological charge density}
and $\mathcal{L}_{eff}$ also contains $Q$ as an auxiliary field.
The correct effective Lagrangian, satisfying the transformation property
\eqref{trasfanomal}, was derived in Refs.
\cite{Witten1980,DV1980,RST1980,KO1980,NA1981} and is given by
\begin{equation}\label{lageff}
\mathcal{L}_2(U,U^{\dagger},Q) = \mathcal{L}_0(U,U^{\dagger})
+\frac{B_{m}}{2\sqrt2}\Tr[MU+M^{\dagger}U^{\dagger}]
+ \frac{i}{2}Q\Tr[\log U-\log U^{\dagger}]+\frac{1}{2A}Q^{2} ,
\end{equation}
where $A=-i\int d^{4}x \langle TQ(x)Q(0)\rangle|_{YM}$ is the so-called
topological susceptibility in the pure Yang--Mills (YM) theory.
After integrating out the variable $Q$ in the effective Lagrangian
\eqref{lageff}, we are left with
\begin{equation}\label{lageffint}
\mathcal{L}_2(U,U^{\dagger})=\mathcal{L}_{0}(U,U^{\dagger})
+\frac{B_{m}}{2\sqrt2}\Tr[MU+M^{\dagger}U^{\dagger}]
+\frac{1}{8}A\left\{\Tr[\log U-\log U^{\dagger}]\right\}^2 ,
\end{equation}
to be compared with Eqs. \eqref{lageffstandard}--\eqref{thooftterm}.

For studying the phase structure of the theory at finite temperature, all
the parameters appearing in the effective Lagrangian must be considered as
functions of the physical temperature $T$. In particular, the parameter
$\rho_\pi$, appearing in the first term of the potential $V_0(U,U^\dag)$
in Eq. \eqref{sigmamodel}, is responsible for the behavior of the theory
across the chiral phase transition at $T=T_{c}$.
Let us consider, for a moment, only the linear sigma model $\lsigma$, i.e.,
let us neglect both the {\it anomalous} symmetry-breaking term and the mass
term in Eq. \eqref{lageffint}.
If $\rho_\pi(T<T_{c}) > 0$, then the value $\ov{U}$ for which the potential
$V_0$ is minimum (that is, in a mean-field approach, the {\it vacuum
expectation value} of the mesonic field $U$) is different from zero and
can be chosen to be
\begin{equation}\label{UWDVminore}
\ov U\vert_{\ropi >0}=v\mathbf{I}, ~~~~
v\equiv\frac{F_\pi}{\sqrt{2}} = \sqrt{\frac{\ropi\lambda_\pi^2}
{\lambda_\pi^2+N_f\lambda_\pi^{\prime 2}}} ,
\end{equation}
which is invariant under the vectorial $U(N_f)_V$ subgroup; the chiral symmetry
is thus spontaneously broken down to $U(N_f)_V$.
Instead, if $\rho_\pi(T>T_{c}) < 0$, we have that
\begin{equation}\label{UWDVmaggiore}
\ov{U}\vert_{\ropi<0}=0 ,
\end{equation}
and the chiral symmetry is realized {\it \`a la} Wigner--Weyl. The critical
temperature $T_{c}$ for the chiral phase transition is thus, in this case,
simply the temperature at which the parameter $\rho_\pi$ vanishes:
$\rho_\pi(T_{c})=0$.

For $T>T_{c}$, where $\rho_\pi < 0$ and $\ov{U}=0$, it is convenient to
use for the matrix field $U$ the simple {\it linear} parametrization
\begin{equation}\label{Ulinear}
U_{ij} = a_{ij} + i b_{ij} ,
\end{equation}
where $a_{ij}$ and $b_{ij}$ are $2N_f^2$ real fields, for which the
vacuum expectation values vanish ($\ov{a}_{ij}=\ov{b}_{ij}=0$).
Inserting Eq. \eqref{Ulinear} into Eq. \eqref{sigmamodel}, and putting
$\rho_\pi \equiv -\frac{1}{2}\bpq$, we find that, up to terms
of second order in the fields,
$\mathcal{L}_0 = \frac{1}{2} \partial_\mu a_{ij} \partial^\mu a_{ij}
+ \frac{1}{2} \partial_\mu b_{ij} \partial^\mu b_{ij}
- \frac{1}{4}\lpq\bpq(a_{ij}^2+b_{ij}^2) + \dots$,
i.e., we have $2N_f^2$ mesonic excitations with equal squared masses
$M^2_U = \frac{1}{2}\lpq\bpq$.

Instead, for $T<T_{c}$, where $\rho_\pi > 0$ and
$\ov{U}=\frac{F_\pi}{\rad2}\mathbf{I}$,
it is more convenient to use for the matrix field $U$ the {\it nonlinear}
parametrization ({\it polar decomposition})
\begin{equation}\label{Unonlinear}
U(x) = H(x)\Gamma(x) = \left(\frac{F_\pi}{\rad2}\mathbf{I}+
\tilde{H}(x)\right) e^{i\frac{\rad2}{F_\pi}\Phi(x)} ,
\end{equation}
where $H = \frac{F_\pi}{\rad2}\mathbf{I}+ \tilde{H}$
is an Hermitian $N_f \times N_f$ matrix, while
$\Gamma = e^{i\frac{\rad2}{F_\pi}\Phi}$ is a unitary $N_f \times N_f$ matrix;
i.e., $\tilde{H}(x) = \frac{1}{\rad2}\sum_{a} h_a(x)\tau^a
+\frac{1}{\rad{N_f}}h_0(x)\mathbf{I}$ and
$\Phi(x) = \frac{1}{\rad2}\sum_{a} \pi_a(x)\tau^a
+\frac{1}{\rad{N_f}}S_\pi(x)\mathbf{I}$
are two Hermitian matrix fields, where $\tau^a$ ($a=1,\dots,N_f^2-1$) are
the generators of the $SU(N_f)$ algebra in the fundamental representation,
with the normalization $\Tr(\tau^a\tau^b)=2\delta_{ab}$
(for $N_f=2$, they are the Pauli matrices, while for $N_f=3$, they are the
Gell-Mann matrices), and $h_a$, $h_0$ are {\it scalar} mesonic fields,
while $\pi_a$, $S_\pi$ are {\it pseudoscalar} mesonic fields, for which the
vacuum expectation values vanish ($\ov{h}_a = \ov{h}_0 = \ov{\pi}_a =
\ov{S}_\pi = 0$).
Inserting Eq. \eqref{Unonlinear} into Eq. \eqref{sigmamodel},
and making use of Eq. \eqref{UWDVminore}, we find that the fields $\pi_a$ and
$S_\pi$ are massless, and they are just the $N_f^2$ (pseudoscalar) Goldstone
bosons generated by the spontaneous breaking of the chiral symmetry down to
the $U(N_f)_V$ subgroup,
while the (scalar) fields $h_a$ ($a=1,\dots,N_f^2-1$) and $h_0$ have
nonzero squared masses, respectively, given by $M^2_a = \lpq\fpq$ and
$M^2_0 = (\lpq+N_f\lambda_\pi^{\prime 2})\fpq$.\footnote{If one is interested,
e.g., at $T=0$, only in the lowest-energy effective
states, i.e., only in the pseudoscalar mesonic excitations, one can formally
decouple the massive scalar excitations $\tilde{H}$, by taking the limit
$\lambda_{\pi}^2 \rightarrow\infty$, which is a ``static,'' i.e.,
infinite-mass, limit for $\tilde{H}$, and thus implies $\tilde{H} \to 0$.
In this limit, the expression \eqref{Unonlinear} for the mesonic field $U$
reduces to $U=\frac{F_\pi}{\rad2}e^{i\frac{\rad2}{F_\pi}\Phi}$, i.e.,
$U U^\dagger = \frac{F^2_\pi}{2}{\bf I}$,
and the effective Lagrangian with this constraint becomes a
{\it nonlinear sigma model}.
We also observe that the quantity $F_\pi$, defined in Eq. \eqref{UWDVminore} as
$F_\pi \equiv \sqrt{2} v$, is just the usual {\it pion decay constant},
since the $SU(N_f)$ axial currents turn out to be, using Eq. \eqref{Unonlinear},
$A^\mu_a = \frac{i}{2} \Tr[T^a \{ \de^\mu U,U^\dag\}]
- \frac{i}{2} \Tr[T^a \{ \de^\mu U^\dag,U\}] = -\sqrt{2} v \de^\mu \pi_a
+ \dots \equiv -F_\pi \de^\mu \pi_a + \dots$.}

If we now take into account the {\it anomalous} term in Eq. \eqref{lageffint}
(while keeping, for simplicity, the chiral limit $M=0$), it is easy to see
that, for $T<T_{c}$, it modifies the result simply by adding
a quadratic term in the pseudoscalar singlet field $S_\pi$,
\begin{equation}\label{lageff2}
\mathcal{L}_2^{(M=0)} = \mathcal{L}_0 - \frac{1}{2} \left(
\frac{2N_f A}{F_\pi^2} \right) S_\pi^2 ,
\end{equation}
from which one derives the famous {\it Witten--Veneziano formula} for
the singlet squared mass (in the chiral limit):
$M^2_{S_\pi} = \frac{2N_f A}{F_\pi^2}$.
However, the {\it anomalous} term in Eq. \eqref{lageffint} makes sense
only in the low-temperature phase ($T<T_{c}$), and it is singular for
$T>T_{c}$, where the vacuum expectation value of the mesonic field $U$
vanishes. On the contrary, the interaction term \eqref{thooftterm}
behaves well both in the low- and high-temperature phases.

\noindent
{\bf A. Effective Lagrangian with the inclusion of a
$U(1)$ axial condensate}

The above-mentioned problems can be overcome by considering a {\it modified}
effective Lagrangian (which was originally proposed in
Refs. \cite{EM1994a,EM1994b,EM1994c} and elaborated on in Refs.
\cite{MM2003,EM2004,EM2011}), which generalizes the Lagrangian
$\mathcal{L}_2$ written in Eq. \eqref{lageff},
so that it correctly satisfies the transformation property \eqref{trasfanomal}
under the chiral group, but also includes an interaction term containing the
determinant of the mesonic field $U$, of the kind of that in
Eq. \eqref{thooftterm}, assuming that there is a
$U(1)_A$-breaking condensate that (possibly) survives across the chiral
transition at $T_{c}$, staying different from zero up to a temperature
$T_{U(1)} > T_{c}$. (Of course, it is also possible that $\tuone \to \infty$,
as a limit case. Another possible limit case, i.e., $\tuone = \tc$, will be
discussed in the concluding comments in Sec. 5.)
The new $U(1)$ chiral condensate has the form
$C_{U(1)} = \langle {\cal O}_{U(1)} \rangle$,
where, for a theory with $N_f$ light quark flavors, ${\cal O}_{U(1)}$ is a
$2N_f$-quark local operator that has the chiral transformation properties of
\cite{tHooft1976,KM1970,Kunihiro2009}
${\cal O}_{U(1)} \sim \displaystyle{{\det_{st}}(\bar{q}_{sR}q_{tL})
+ {\det_{st}}(\bar{q}_{sL}q_{tR}) }$,
where $s,t = 1, \dots ,N_f$ are flavor indices. The color indices (not
explicitly indicated) are arranged in such a way that
(i) ${\cal O}_{U(1)}$ is a color singlet, and (ii)
$C_{U(1)} = \langle {\cal O}_{U(1)} \rangle$ is a {\it genuine} $2N_f$-quark
condensate, i.e., it has no {\it disconnected} part proportional to some
power of the quark-antiquark chiral condensate $\langle \bar{q} q \rangle$;
the explicit form of the condensate for the cases $N_f=2$ and $N_f=3$ is
discussed in detail in the Appendix A of Ref. \cite{EM2011} (see also Refs.
\cite{EM1994c,DM1995}).

The modified effective Lagrangian is written in terms of the topological charge
density $Q$, the mesonic field $U_{ij} \sim \bar{q}_{jR} q_{iL}$ (up to a
multiplicative constant), and the new field variable $X \sim {\det} \left(
\bar{q}_{sR} q_{tL} \right)$ (up to a multiplicative constant), associated
with the $U(1)$ axial condensate \cite{EM1994a,EM1994b,EM1994c},
\begin{eqnarray}\label{lagtot}
\lefteqn{
\La (U,U^\dagger ,X,X^\dagger ,Q)
= \frac{1}{2}\Tr[\partial_\mu U\partial^\mu U^\dagger]
+ \frac{1}{2}\partial_\mu X\partial^\mu X^\dagger } \nonumber \\
& & -V(U,U^\dagger ,X,X^\dagger)
+ \frac{i}{2}\omega_1 Q \Tr[\log U - \log U^\dagger] \nonumber \\
& & + \frac{i}{2}(1-\omega_1)Q[\log X-\log X^\dagger] + \frac{1}{2A}Q^2,
\end{eqnarray}
where the potential term $V(U,U^{\dagger},X,X^{\dagger})$ has the form
\begin{eqnarray}\label{V}
V(U,U^{\dagger},X,X^{\dagger}) &=&
\frac{1}{4}\lambda_{\pi}^{2}\Tr[(UU^{\dagger}-\rho_{\pi}{\bf I})^{2}]
+\frac{1}{4}\lambda_{\pi}^{\prime 2}[\Tr(UU^{\dagger})]^{2}
+\frac{1}{4}\lambda_{X}^{2}[XX^{\dagger}-\rho_{X}]^{2} \nonumber \\
&-&\frac{B_{m}}{2\sqrt{2}}\Tr[MU+M^{\dagger}U^{\dagger}]
-\frac{c_{1}}{2\sqrt{2}}[X^{\dagger}\det U+X\det U^{\dagger}].
\end{eqnarray}
Since under chiral $U(N_f)_L\otimes U(N_f)_R$ transformations [see Eq.
\eqref{trasfU}] the field $X$ transforms exactly as $\det U$,
\begin{equation}\label{trasfX}
U(N_f)_L\otimes U(N_f)_R:~~~~ X \rightarrow \det(V_L)\det(V_R)^* X,
\end{equation}
[i.e., $X$ is invariant under $SU(N_f)_L\otimes SU(N_f)_R\otimes U(1)_V$,
while, under a $U(1)$ axial transformation \eqref{U1trasf},
$X\rightarrow e^{-i2N_f\alpha}X$],
we have that, in the chiral limit $M=0$, the effective Lagrangian
\eqref{lagtot} is invariant under $SU(N_f)_L\otimes SU(N_f)_R \otimes U(1)_V$,
while under a $U(1)$ axial transformation, it correctly transforms as
in Eq. \eqref{trasfanomal}.

After integrating out the variable $Q$ in the effective Lagrangian
\eqref{lagtot}, we are left with
\begin{equation}\label{lagtotb}
\mathcal{L}(U,U^{\dagger},X,X^{\dagger})
= \frac{1}{2}\Tr[\partial_{\mu}U\partial^{\mu}U^{\dagger}]
+ \frac{1}{2}\partial_{\mu}X\partial^{\mu}X^{\dagger}
- \tilde{V}(U,U^{\dagger},X,X^{\dagger}) ,
\end{equation}
where
\begin{eqnarray}\label{Vtilde}
\tilde{V}(U,U^{\dagger},X,X^{\dagger}) &=& V(U,U^{\dagger},X,X^{\dagger})
\nonumber \\
&-& \frac{1}{8}A\{\omega_{1}\Tr[\log U-\log U^{\dagger}]
+ (1-\omega_{1})[\log X-\log X^{\dagger}]\}^2 .
\end{eqnarray}
As we have already said, all the parameters appearing in the effective
Lagrangian must be considered as functions of the physical temperature $T$.
In particular, the parameters $\rho_{\pi}$ and $\rho_X$ determine the
expectation values $\langle U \rangle$ and $\langle X \rangle$, and so they
are responsible for the behavior of the theory across the
$SU(N_f) \otimes SU(N_f)$ and the $U(1)$ chiral phase transitions.
We shall assume that the parameters $\rho_\pi$ and $\rho_X$, as functions
of the temperature $T$, behave as reported in Table \ref{table1};
$\tro$ is thus the temperature at which the parameter $\ropi$ vanishes, while
$\tuone>\tro$ is the temperature at which the parameter $\rox$ vanishes
(with, as we have said above, $\tuone\to\infty$, i.e., $\rox>0$ $\forall T$,
as a possible limit case).
\begin{table}[htbp]
\begin{center}
\begin{tabular}{|l|c|c|r|}
\hline
$T<\tro$ & $\tro<T<T_{U(1)}$ & $T>T_{U(1)}$ \\ 
\hline
$\ropi>0$ & $\ropi<0$ & $\ropi<0$ \\ 
\hline
$\rox>0$ & $\rox>0$ & $\rox<0$ \\ 
\hline
\end{tabular}
\end{center}
\caption{Dependence of the parameters $\ropi$ and $\rox$ on the
temperature $T$.}
\label{table1}
\end{table}
We shall see in the next section that, in the case $N_f\geq 3$, one has
$\tc=\tro$ (exactly as in the case of the linear sigma model $\mathcal{L}_0$
discussed above), while, as we shall see in Sec. 4, the situation in which
$N_f=2$ is more complicated, being $\tro<\tc<\tuone$ in that case (unless
$\tro=\tc=\tuone$; this limit case will be discussed in the concluding
comments in Sec. 5).

Concerning the parameter $\omega_{1}$, in order to avoid a singular behavior of
the anomalous term in Eq. \eqref{Vtilde} above the chiral transition temperature
$T_{c}$, where the vacuum expectation value of the mesonic field $U$
vanishes (in the chiral limit $M=0$), we shall assume that
$\omega_{1}(T\geq T_{c})=0$.

Finally, let us observe that the interaction term between the
$U$ and $X$ fields in Eq. \eqref{V}, i.e.,
\begin{equation}\label{Lint}
\mathcal{L}_{int}=\frac{c_{1}}{2\rad2}[X^{\dagger}\det U+X\det U^{\dagger}] ,
\end{equation}
is very similar to the interaction term \eqref{thooftterm}
that we have discussed above for the effective Lagrangian $\mathcal{L}_1$.
However, the term \eqref{Lint} is not {\it anomalous}, being invariant
under the chiral group $U(N_f)_L \otimes U(N_f)_R$, by virtue of Eqs.
\eqref{trasfU} and \eqref{trasfX}.
Nevertheless, if the field $X$ has a (real) {\it nonzero} vacuum expectation
value $\ov X$ [the $U(1)$ axial condensate], then we can write
\begin{equation}\label{hXSx}
X=(\ov{X}+h_X)e^{i\frac{S_X}{\ov{X}}} ~~~~
({\rm with:}~~ \ov{h}_X = \ov{S}_X = 0),
\end{equation}
and, after susbstituting this in Eq.  \eqref{Lint} and expanding in powers of
the excitations $h_X$ and $S_X$, one recovers, at the leading order, an
interaction term of the form \eqref{thooftterm}:
\begin{equation}\label{Lint-expanded}
\mathcal{L}_{int} = c_{I}[\det U+\det U^{\dagger}] + \dots ,~~~~{\rm with:}~~
c_{I}=\frac{c_1\ov X}{2\rad2}.
\end{equation}
In the rest of this paper we shall analyze in detail the effects of assuming
a nonzero value of the $U(1)$ axial condensate $\ov{X}$ on the scalar and
pseudoscalar meson mass spectrum {\it above} the chiral transition temperature
($T>T_{c}$), both for the case $N_f\geq 3$ (Sec. 3) and for the case $N_f=2$
(Sec. 4).

\newsection{Mass spectrum for $T>T_{c}$ in the case $N_f\geq 3$}

\noindent
The results for the scalar and pseudoscalar meson mass spectrum for $T > T_{c}$
in the case $N_f\geq 3$ were rapidly sketched in Ref. \cite{EM1994a} and, in
this section, we shall rederive them in a more detailed and accurate way
in order to allow for a more clear comparison with the novel results that we
shall obtain in the next section for the case $N_f=2$.

Let us suppose to be in the range of temperatures $\tro<T<\tuone$, where,
according to Table \ref{table1},
\begin{equation}\label{Trange}
\ropi\equiv-\frac{1}{2}B_\pi^2<0,~~~~ \rox\equiv\frac{1}{2}F_X^2>0 .
\end{equation}
Since we expect that, due to the sign of the parameter $\rho_X$ in the
potential \eqref{V}, the $U(1)$ axial symmetry is broken by a nonzero
vacuum expectation value of the field $X$ (at least for $\lxq \to \infty$
we should have $\ov X^\dag \ov X \to \frac{1}{2} F_X^2$),
we shall use for the field $U$ the linear parametrization \eqref{Ulinear},
while for the field $X$, we shall use a nonlinear parametrization,
similar to the polar decomposition in Eq. \eqref{Unonlinear},
\begin{equation}\label{paramUXL3}
U_{ij}=a_{ij}+ib_{ij}, \,\,\,\,X=\alpha e^{i\beta}=(\ov{\alpha}+h_X)
e^{i\left(\ov{\beta}+\frac{S_X}{\ov{\alpha}}\right)},
\end{equation}
where $\ov{X}=\ov{\alpha}e^{i\ov{\beta}}$ (with $\ov\alpha \neq 0$)
is the vacuum expectation value of $X$ and $a_{ij}$, $b_{ij}$, $h_X$, and
$S_X$ are real fields.
Inserting Eq. \eqref{paramUXL3} into the expressions \eqref{V} and
\eqref{Vtilde}, we find the expressions for the potential with and without the
anomalous term (with $\omega_1=0$),
\begin{equation}\label{Vanom}
\tilde{V} = V-\frac{1}{8}A[\log X-\log X^{\dagger}]^{2}
= V+\frac{1}{2}A\beta^2
\end{equation}
and
\begin{equation}\label{VL3}
\begin{split}
V=&~\frac{N_f}{16}\lpq B_\pi^4 +\frac{1}{4}\lpq \Tr[(UU^\dag)(UU^\dag)]
+\frac{1}{4}\lambda_{\pi}^{\prime 2}[\Tr(UU^{\dagger})]^{2}
+\frac{1}{4}\lxq\left(\alpha^2-\frac{1}{2}F_X^2\right)^{2} \\
&+\frac{1}{4}\lpq\bpq(a_{ij}^2+b_{ij}^2)
-\frac{B_m}{\rad2}(m_{ij}a_{ji}-n_{ij}b_{ji}) \\
&-\frac{c_{1}}{2\rad2}[\alpha\cos\beta(\det U+\det U^\dag)+i\alpha\sin\beta(\det U-\det U^\dag)] ,
\end{split}
\end{equation}
where we have assumed the most general (complex) mass matrix
$M_{ij}=m_{ij}+in_{ij}$, with $m_{ij}$ and $n_{ij}$ real.
Let us first look for the equations for a stationary point ($S$) of the
nonanomalous potential $V$, indicating with $\ov{U}$ and $\ov{X}$
the values of the fields $U$ and $X$ in this point:
\begin{equation}\label{eqminL3}
\begin{matrix}
&\frac{\de V}{\de a_{ij}}\vert_{S}=\frac{1}{2}\lpq\bpq\ov{a}_{ij}-\frac{B_m}{\rad 2}m_{ji} + \dots =0 , \\ \\
&\frac{\de V}{\de b_{ij}}\vert_S=\frac{1}{2}\lpq\bpq\ov{b}_{ij}+\frac{B_m}{\rad 2}n_{ji} + \dots =0 , \\ \\
&\frac{\de V}{\de \alpha}\vert_S=\lambda_{X}^{2}
\left(\ov\alpha^{2}-\frac{F_{X}^{2}}{2}\right)\ov\alpha-\frac{c_{1}}{2\sqrt2}
[\cos\ov\beta(\det\ov U+\det\ov U^{\dagger})\\ 
&+i\sin\ov\beta(\det\ov U^{\dagger}-\det\ov U)] =0 , \\ \\
&\frac{\de V}{\de \beta}\vert_S=\frac{c_{1}}{2\sqrt2}\ov\alpha 
[\sin\ov\beta(\det\ov U + \det\ov U^{\dagger})
-i\cos\ov\beta(\det\ov U^{\dagger}- \det\ov U)]=0 . 
\end{matrix}
\end{equation}
From the first two equations, where we have omitted terms that, for $N_f\geq 3$,
are of order 2 or higher in the fields $a$ and $b$, we find that, at the
leading order in $M$,
\begin{equation}\label{minUL3}
\ov{U}=\frac{2B_{m}}{\sqrt{2}\lambda_{\pi}^{2}B_{\pi}^{2}}M^\dag+\dots.
\end{equation}
Let us now consider the second derivatives of the potential $V$ with respect
to the fields, calculated at the stationary point $S$:
\begin{equation}\label{HessianL3}
\begin{matrix}
\frac{\de^2 V}{\de a_{lm} \de a_{ij}}\vert_S=\frac{1}{2}\lpq\bpq\delta_{il}\delta_{jm} +\dots , \\ \\
\frac{\de^2 V}{\de b_{lm} \de b_{ij}}\vert_S=\frac{1}{2}\lpq\bpq\delta_{il}\delta_{jm} +\dots , \\ \\
\frac{\de^2 V}{\de \alpha^2}\vert_S=\lxq\left(3\ov\alpha^2-\frac{F_X^2}{2}\right) , \\ \\
\frac{\de^2 V}{\de \beta^2}\vert_S=\frac{c_{1}}{2\sqrt2}\ov\alpha 
[\cos\ov\beta(\det\ov U + \det\ov U^{\dagger})
+i\sin\ov\beta(\det\ov U^{\dagger}- \det\ov U)] , \\ \\
\frac{\de^2 V}{\de \alpha \de \beta}\vert_S=\frac{c_{1}}{2\sqrt2}
[\sin\ov\beta(\det\ov U + \det\ov U^{\dagger})
-i\cos\ov\beta(\det\ov U^{\dagger}- \det\ov U)] .
\end{matrix}
\end{equation}
The first two equations are given at the leading order in the quark masses,
and all the second derivatives, which are not shown in Eq. \eqref{HessianL3},
are of order $\Ord(m)$ or higher in the quark masses. From the third equation
of Eqs. \eqref{HessianL3}, it is clear that the stationary point can be a
{\it minimum} of the potential only for $\ov\alpha \neq 0$.
If we now take for $M$ the {\it physical} real diagonal matrix
$M = {\rm diag}(m_1,\dots,m_{N_f})$, we have that $M=M^\dag$ and therefore,
by virtue of the result \eqref{minUL3}, also $\ov U=\ov U^\dag$.
Indeed, this is a more general result, not directly related to the particular
solution \eqref{minUL3} (which, as we shall see in the next section, is valid
for $N_f\geq 3$, but not for $N_f=2$), being due, when $M$ is a real
diagonal matrix (or, more generally, when $M$ is Hermitian), to the invariance
of the theory under {\it parity} ($P$) transformations [i.e., being $U_{ij} \sim
\bar{q}_{jR} q_{iL}$ and $X \sim {\det} \left( \bar{q}_{sR} q_{tL} \right)$,
$U(x^0,\vec{x}) \stackrel{P}{\rightarrow}U^\dag(x^0,-\vec{x})$,
$X(x^0,\vec{x})\stackrel{P}{\rightarrow}X^\dag(x^0,-\vec{x})$],
which requires that $\ov{U}=\ov{U}^\dag$ and $\ov{X}=\ov{X}^\dag$.
From the last of Eqs. \eqref{eqminL3}, we thus find that
$\sin\ov{\beta}=0$, i.e., $\ov\beta=0,\pi$, which also implies that
$\frac{\de^2 V}{\de \alpha \de \beta}\vert_S=0$.
Moreover, from the fourth Eq. \eqref{HessianL3}, using the result
\eqref{minUL3}, it is clear that, for the stationary point $S$ to be a minimum,
we must require, assuming $c_1>0$ and $B_m>0$, that also
$\ov\alpha\cos\ov\beta > 0$; so, finally, we can take $\ov\alpha > 0$
and $\ov\beta = 0$.
We can then determine $\ov\alpha$ using the third equation in Eqs.
\eqref{eqminL3} and so find
\begin{equation}\label{alphaminL3sol}
\ov\alpha = \frac{F_X}{\rad2} + \frac{c_1}{\rad2\lxq\fxq}\left(
\frac{2B_m}{\rad2 \lpq\bpq}\right)^{N_f} \det M + \dots ,
\end{equation}
which gives
$\frac{\de^2 V}{\de \alpha^2}\vert_S=\lxq F_X^2 + \mathcal{O}(\det M)$.

If we now consider the {\it full} potential $\tilde V$, with the inclusion
of the anomalous term, see Eq. \eqref{Vanom}, it is trivial to see that the
solution that we have found for the minimum of $V$, given by Eqs.
\eqref{minUL3} and \eqref{alphaminL3sol} with $\ov\beta=0$, is also a
minimum for the potential $\tilde V$, the only modification being in the
second derivative of the potential with respect to $\beta$, which
is now given by [see the fourth equation in Eqs. \eqref{HessianL3}]
$\frac{\de^2 \tilde V}{\de \beta^2}\vert_S
= \frac{\de^2 V}{\de \beta^2}\vert_S + A = A + \mathcal{O}(\det M)$.

In particular, in the chiral limit $M=0$, we find that
$\ov U = 0$ and $\ov X = \ov\alpha = \frac{F_X}{\rad2}$,
which means that, in this range of temperatures $\tro<T<\tuone$, the
$SU(N_f)_L\otimes SU(N_f)_R$ chiral symmetry is restored so that we can say
that (at least for $N_f\geq 3$) $\tc\equiv\tro$, while the $U(1)$ axial
symmetry is broken by the $U(1)$ axial condensate $\ov X$.
Concerning the mass spectrum of the effective Lagrangian, we have $2N_f^2$
degenerate scalar and pseudoscalar mesonic excitations, described by the fields
$a_{ij}$ and $b_{ij}$, {\it plus} a scalar singlet field
$h_X = \alpha - \ov\alpha$ and a pseudoscalar singlet field
$S_X = \ov\alpha \beta$ [see Eq. \eqref{paramUXL3}],
with squared masses given by
\begin{equation}\label{massesL3}
M^2_U = \frac{1}{2}\lpq\bpq,~~~~ M^2_{h_X}=\lxq\fxq,~~~~
M^2_{S_X} = \frac{A}{\ov\alpha^2} = \frac{2A}{\fxq}.
\end{equation}
While the mesonic excitations described by the field $U$ are of the usual
$q \bar q$ type, the scalar singlet field $h_X$ and the pseudoscalar singlet
field $S_X$ describe instead two {\it exotic}, $2N_f$-quark excitations
of the form $h_X \sim {\det}(\bar{q}_{sL}q_{tR}) + {\det}(\bar{q}_{sR}q_{tL})$
and $S_X \sim i[ {\det}(\bar{q}_{sL}q_{tR}) - {\det}(\bar{q}_{sR}q_{tL})]$.
In particular, the physical interpretation of the pseudoscalar singlet
excitation $S_X$ is rather obvious, and it was already discussed in Ref.
\cite{EM1994a}: it is nothing but the {\it would-be} Goldstone particle
coming from the breaking of the $U(1)$ axial symmetry. In fact, neglecting 
the anomaly, it has zero mass in the chiral limit of zero quark masses.
Yet, considering the anomaly, it acquires a {\it topological} squared
mass proportional to the topological susceptibility $A$ of the pure YM theory,
as required by the Witten--Veneziano mechanism \cite{Witten1979,Veneziano1979}.

\newsection{Mass spectrum for $T>\tc$ in the case $N_f=2$}

\noindent
In this section we shall derive the results for the scalar and pseudoscalar
mesonic mass spectrum for $T > T_{c}$ in the case $N_f=2$, with a quark
mass matrix given by $M = {\rm diag}(m_u,m_d)$.

As in the previous section, we start considering the range of temperatures
$\tro<T<\tuone$, with the parameters $\rho_\pi$ and $\rho_X$ given by
Eq. \eqref{Trange} (see also Table \ref{table1}).
We shall use for the field $U$ a more convenient variant of the linear
parametrization \eqref{Ulinear}, while for the field $X$, we shall use the
usual nonlinear parametrization given in Eq. \eqref{paramUXL3},\footnote{Here,
we immediately put $\ov\beta=0$, since, as one can easily see, the
arguments leading to $\ov\beta=0$, which we have given in the previous section,
are valid also for $N_f=2$.}
\begin{equation}\label{parUX}
U = \frac{1}{\sqrt2} [ (\sigma+i\eta)\mathbf{I}
+ (\vec{\delta}+i\vec{\pi})\cdot\vec{\tau} ] ,~~~~
X = \alpha e^{i\beta} = (\ov{\alpha}+h_X) e^{i\frac{S_X}{\ov{\alpha}}},
\end{equation}
where $\tau^a$ ($a=1,2,3$) are the three Pauli matrices [with the usual
normalization $\Tr(\tau^a\tau^b)=2\delta_{ab}$] and the multiplicative factor
$\frac{1}{\rad2}$ guarantees the correct normalization of the kinetic term
in the effective Lagrangian. The fields $\sigma$, $\eta$, $\vec\delta$, and
$\vec\pi$ describe, respectively, the isoscalar ($I=0$) scalar $q\bar q$ mesonic
excitation $\sigma$ (also known as $f_0$ in the modern language of hadron
spectroscopy), the isoscalar ($I=0$) pseudoscalar $q\bar q$ mesonic excitation
$\eta$, the isovector ($I=1$) scalar $q\bar q$ mesonic excitation $\vec\delta$
(also known as $\vec{a}_0$), and the isovector ($I=1$) pseudoscalar $q\bar q$
mesonic excitation $\vec\pi$.

Inserting Eq. \eqref{parUX} into the expressions \eqref{V} and \eqref{Vtilde},
we find the following expression for the potential $V$, without the
anomalous term,
\begin{equation}\label{VL2}
\begin{split}
V=&~\frac{1}{8}\lpq B_{\pi}^4+\frac{1}{8}\Lambda_{\pi}^{2}(\sigma^{2}+\eta^{2}+\vec{\pi}^{2}+\vec{\delta}^{2})^2
+\frac{1}{2}\lambda_{\pi}^{2}(\sigma^{2}\vec{\delta}^{2}+2\sigma\eta\vec{\delta}\cdot\vec{\pi}+\eta^{2}\vec{\pi}^{2})\\
&+\frac{1}{2}\lambda_{\pi}^{2}[\vec{\pi}^{2}\vec{\delta}^{2}-(\vec{\pi}\cdot\vec{\delta})^{2}]+
\frac{1}{4}\lambda_{\pi}^{2}B_{\pi}^{2}[\sigma^{2}+\eta^{2}+\vec{\delta}^{2}+\vec{\pi}^{2}]+
\frac{1}{4}\lambda_{X}^{2}\left(\alpha^{2}-\frac{F_{X}^{2}}{2}\right)^{2}\\
&-\frac{B_{m}}{2}[(m_{u}+m_{d})\sigma+(m_{u}-m_{d})\delta_{3}] \\
&-\frac{c_{1}}{2\sqrt2}[\alpha \cos\beta(\sigma^{2}-\eta^{2}-\vec{\delta}^{2}+\vec{\pi}^{2})+
2\alpha \sin\beta(\sigma\eta-\vec{\delta}\cdot\vec{\pi})],
\end{split}
\end{equation}
where
\begin{equation}\label{definitionLambda}
\Lambda_{\pi}^{2}\equiv\lambda_{\pi}^{2}+2\lambda_{\pi}^{\prime 2},
\end{equation}
while the full potential $\tilde V$, including also the anomalous term
(with $\omega_1=0$), is still given by Eq. \eqref{Vanom}, i.e.,
$\tilde{V} = V+\frac{1}{2}A\beta^2$.

When looking for the equations for a stationary point ($S$) of the
potential $\tilde{V}$, indicating as usual with $\ov{U}$ and $\ov{X}$
the values of the fields $U$ and $X$ in this point,
we can immediately make use, with $M$ being a real diagonal (and, therefore,
Hermitian) matrix, of the invariance of the theory under parity ($P$)
transformations (as already observed in the previous section),
which requires that $\ov{U}=\ov{U}^\dag$ and $\ov{X}=\ov{X}^\dag$.
That is to say, using the parametrization \eqref{parUX},
$\ov\eta = \ov\pi_a = \ov\beta = 0$. This automatically guarantees the
vanishing of the first derivatives of the potential $\tilde{V}$ with
respect to the pseudoscalar fields at the stationary point $S$, i.e.,
$\frac{\partial\tilde{V}}{\partial\eta}\vert_S =
\frac{\partial\tilde{V}}{\partial\pi_a}\vert_S =
\frac{\partial\tilde{V}}{\partial\beta}\vert_S = 0$,
as one can easily verify using Eqs. \eqref{VL2} and \eqref{Vanom}. 

Moreover, the vanishing, at the stationary point ($S$), of the derivatives of
Eq. \eqref{VL2} with respect to the fields $\delta_a$ ($a=1,2,3$),
gives the following three equations:
\begin{equation}\label{dedelta}
\frac{\partial\tilde{V}}{\partial\delta_a}\vert_S =
\frac{1}{2} \left[ \Lambda_{\pi}^{2}(\ov\sigma^{2} + \ov{\vec\delta}^{2})
+ 2\lambda_{\pi}^{2}\ov\sigma^{2} + (\lambda_{\pi}^{2}B_{\pi}^{2}
+ \sqrt{2}c_1\ov\alpha) \right] \ov\delta_a
- \frac{1}{2}B_m (m_u-m_d) \delta_{a3}=0.
\end{equation}
For $a=1$ and $a=2$, using the fact that $c_1>0$ and $\ov\alpha>0$ (or,
more generally, $c_1\ov\alpha>0$; see the discussion in the previous section,
which can be easily extended also to the case $N_f=2$ considered here),
one immediately finds the solution $\ov\delta_1 = \ov\delta_2 = 0$.
Let us also observe that, in the chiral limit $m_u=m_d=0$, or, more
generally, in the limit of equal quark masses $m_u=m_d$, one also has
$\ov\delta_3 = 0$ so that $\ov{U} = \frac{\ov\sigma}{\rad2}{\bf I}$,
which is invariant under the $SU(2)_V$ ({\it isospin}) symmetry,
as it must be.

So, finally, we are left with the following three equations for the
values $\ov\alpha$, $\ov\sigma$ and $\ov\delta \equiv \ov\delta_{3}$:
\begin{eqnarray}\label{minUalpha}
\frac{\partial\tilde{V}}{\partial\sigma}\vert_S &=&
\frac{1}{2} \left[ \Lambda_{\pi}^{2}(\ov\sigma^{2} + \ov\delta^{2})
+ 2\lambda_{\pi}^{2}\ov\delta^{2} + (\lambda_{\pi}^{2}B_{\pi}^{2}
- \sqrt{2}c_1\ov\alpha) \right] \ov\sigma - \frac{1}{2}B_m (m_u+m_d)=0,
\nonumber \\
\frac{\partial\tilde{V}}{\partial\delta_3}\vert_S &=&
\frac{1}{2} \left[ \Lambda_{\pi}^{2}(\ov\sigma^{2} + \ov\delta^{2})
+ 2\lambda_{\pi}^{2}\ov\sigma^{2} + (\lambda_{\pi}^{2}B_{\pi}^{2}
+ \sqrt{2}c_1\ov\alpha) \right] \ov\delta - \frac{1}{2}B_m (m_u-m_d)=0,
\nonumber \\
\frac{\partial\tilde{V}}{\partial\alpha}\vert_S &=&
\lambda_{X}^{2}\left(\ov\alpha^{2}-\frac{F_{X}^{2}}{2}\right)\ov\alpha
-\frac{c_{1}}{2\sqrt2}(\ov\sigma^2-\ov\delta^2)=0.
\end{eqnarray}
It is easy to see that Eqs. \eqref{minUalpha} admit the following
solution (at the first nontrivial order in the quark masses $m_u$ and $m_d$),
\begin{eqnarray}\label{solution1}
\smin &=& \frac{B_m}{\lpq\bpq-c_1F_X} (m_u+m_d)+\dots, \nonumber \\
\dmin &=& \frac{B_m}{\lpq\bpq+c_1F_X} (m_u-m_d)+\dots, \nonumber \\
\overline{\alpha} &=& \frac{F_{X}}{\sqrt2} + \frac{\sqrt{2}c_{1}^2\lpq\bpq}
{\lxq F_X(\lambda_{\pi}^{4}B_{\pi}^{4}-c_{1}^{2}F_{X}^{2})^2}
B_m^2 (m_{u}^{2}+m_{d}^{2}) \nonumber \\
&+& \frac{\rad2 c_1 (\lambda_{\pi}^{4}B_{\pi}^{4}+c_{1}^{2}F_{X}^{2})}
{\lxq\fxq(\lambda_{\pi}^{4}B_{\pi}^{4}-c_{1}^{2}F_{X}^{2})^2} B_m^2 m_u m_d
+ \dots ,
\end{eqnarray}
which, in the chiral limit $m_u=m_d=0$, reduces to
\begin{equation}\label{solution1chiral}
\ov{\sigma} = \ov{\delta} = 0,~~~~ \ov{\alpha} = \frac{F_X}{\rad2} ,~~~~
{\rm i.e.:}~~~~ \ov U = 0,~~~~ \ov X = \ov\alpha = \frac{F_X}{\rad2} ,
\end{equation}
signalling that the $SU(N_f)_L\otimes SU(N_f)_R$ chiral symmetry is restored,
while the $U(1)$ axial symmetry is broken by the $U(1)$ axial condensate
$\ov X$.

To see if this stationary point is a minimum of the potential (and,
eventually, in order to derive the mass spectrum of the effective Lagrangian),
we must study the matrix of the second derivatives ({\it Hessian}) of the
potential $\tilde{V}$ with respect to the fields at the stationary point $S$.
By virtue of the parity invariance of the theory, one immediately
has that the {\it mixed} second derivatives of $\tilde{V}$ with respect to
a scalar field and a pseudoscalar field vanish at the stationary point $S$,
as one can easily verify using Eqs. \eqref{VL2} and \eqref{Vanom}.
In other words, the {\it scalar sector} $(h_X,\sigma,\vec\delta)$ and the
{\it pseudoscalar sector} $(S_X,\eta,\vec\pi)$ are decoupled in the
matrix of the second derivatives of $\tilde{V}$ at the stationary point $S$,
and, therefore, they can be studied separately.

\noindent
{\bf A. Scalar sector}

From Eqs. \eqref{VL2} and \eqref{Vanom}, it comes out that the Hessian
matrix (evaluated at the stationary point $S$) is already diagonal
with respect to the fields $\delta_1$ and $\delta_2$, with a common value
of the squared masses given by
\begin{equation}\label{massdelta12}
M_{\delta_{1,2}}^2 = \frac{1}{2}(\lambda_{\pi}^{2}B_{\pi}^{2}
+ c_{1}\sqrt{2}\ov\alpha) + \frac{1}{2}\Lpq(\ov\sigma^2+\ov\delta^2)
+ \lpq\ov\sigma^2 .
\end{equation}
The Hessian of the remaining scalar fields $(h_X, \sigma, \delta_3)$
turns out to be
\begin{equation}\label{HessianSL2}
\mathcal{H}_{(S)} = \left(
\begin{matrix}
\lxq\left(3\ov\alpha^2-\frac{\fxq}{2}\right) & -\frac{c_{1}}{\rad 2}\smin &
\frac{c_{1}}{\rad 2}\dmin \\
-\frac{c_{1}}{\rad 2}\smin & \frac{1}{2}(\lpq\bpq-c_{1}\sqrt{2}\ov\alpha)
+ \Delta_\sigma & (\Lpq+2\lpq)\smin\dmin \\
\frac{c_{1}}{\rad 2}\dmin & (\Lpq+2\lpq)\smin\dmin &
\frac{1}{2}(\lpq\bpq+c_{1}\sqrt{2}\ov\alpha) + \Delta_\delta
\end{matrix}
\right),
\end{equation}
where $\Delta_\sigma \equiv \frac{3}{2}\Lpq\smin^2 +
\frac{1}{2}(\Lpq+2\lpq)\dmin^2$ and
$\Delta_\delta \equiv \frac{1}{2}(\Lpq+2\lpq)\smin^2
+ \frac{3}{2}\Lpq\dmin^2$.
Therefore, in the chiral limit $m_u=m_d=0$, see Eq. \eqref{solution1chiral},
the Hessian matrix of the scalar fields $(h_X,\sigma,\vec\delta)$
turns out to be diagonal, with squared masses given by
\begin{equation}\label{Smasses}
M_{h_X}^2=\lxq\fxq ,~~~~
M_{\sigma}^{2}=\frac{1}{2}(\lpq\bpq - c_{1}F_{X}) ,~~~~
M_{\delta}^{2}=\frac{1}{2}(\lpq\bpq + c_{1}F_{X}) .
\end{equation}

\noindent
{\bf B. Pseudoscalar sector}

From Eqs. \eqref{VL2} and \eqref{Vanom}, it comes out that the Hessian
matrix (evaluated at the stationary point $S$) is already diagonal
with respect to the fields $\pi_1$ and $\pi_2$, with a common value
of the squared masses given by
\begin{equation}\label{masspi12}
M_{\pi_{1,2}}^{2} = \frac{1}{2}(\lambda_{\pi}^{2}B_{\pi}^{2}
- c_{1}\sqrt{2}\ov\alpha) + \frac{1}{2}\Lpq(\ov\sigma^2+\ov\delta^2)
+ \lpq\ov\delta^2 .
\end{equation}
The Hessian of the remaining pseudoscalar fields $(S_X, \eta, \pi_3)$
turns out to be
\begin{equation}\label{HessianPSL2}
\mathcal{H}_{(PS)} = \left(
\begin{matrix}
\frac{c_{1}}{2\sqrt{2}\ov\alpha}(\ov{\sigma}^{2}-\ov{\delta}^{2})
+ \frac{A}{\ov\alpha^2} & -\frac{c_{1}}{\rad 2}\ov{\sigma} &
\frac{c_{1}}{\rad 2}\ov{\delta} \\ 
-\frac{c_{1}}{\rad 2}\ov{\sigma} & \frac{1}{2}(\lpq\bpq+c_{1}\sqrt{2}\ov\alpha)
+ \Delta & \lpq\smin\dmin \\ 
\frac{c_{1}}{\sqrt 2}\dmin & \lpq\dmin\smin &
\frac{1}{2}(\lpq\bpq-c_{1}\sqrt{2}\ov\alpha) + \Delta
\end{matrix}
\right) ,
\end{equation}
where $\Delta \equiv \frac{1}{2}\Lpq(\ov\sigma^2+\ov\delta^2)$.
Therefore, in the chiral limit $m_u=m_d=0$, see Eq. \eqref{solution1chiral},
the Hessian matrix of the pseudoscalar fields $(S_X,\eta,\vec\pi)$
turns out to be diagonal, with squared masses given by
\begin{equation}\label{PSmasses}
M_{S_X}^2=\frac{2A}{\fxq} ,~~~~
M_{\eta}^2=\frac{1}{2}(\lpq\bpq+c_1 F_X) ,~~~~
M_{\pi}^2=\frac{1}{2}(\lpq\bpq-c_1 F_X) .
\end{equation}
Therefore, in the case $N_f=2$, the restoration of the $SU(2)_L \otimes SU(2)_R$
chiral symmetry manifests itself in the appearance, in the mass spectrum
of the effective Lagrangian, of two $q\bar q$ chiral multiplets
$(\frac{1}{2},\frac{1}{2})$, namely,
\begin{eqnarray}\label{multiplets}
(\sigma,\vec\pi)&:&~~~~
M_{\sigma}^{2} = M_{\pi}^{2} = \frac{1}{2}(\lpq\bpq-c_{1}F_{X}), \nonumber \\
(\eta,\vec\delta)&:&~~~~
M_{\eta}^{2} = M_{\delta}^{2} = \frac{1}{2}(\lpq\bpq+c_{1}F_{X}).
\end{eqnarray}
Instead, the squared masses of the $q\bar q$ mesonic excitations belonging
to a same $U(1)$ chiral multiplet, such as $(\sigma,\eta)$ and
$(\vec\pi,\vec\delta)$, remain {\it split} by the quantity
\begin{equation}\label{Msplit}
\Delta M_{U(1)}^2 \equiv M_{\eta}^2-M_{\sigma}^2
= M_{\delta}^{2}-M_{\pi}^{2} = c_1 F_X,
\end{equation}
proportional to the $U(1)$ axial condensate.
This result is to be contrasted with the corresponding result obtained
in the previous section for the $N_f\geq 3$ case, see Eq. \eqref{massesL3},
in which {\it all} (scalar and pseudoscalar) $q\bar q$ mesonic excitations
(described by the field $U$) turned out to be degenerate, with squared masses
$M_U^{2}=\frac{1}{2}\lpq\bpq$.

We must now make an important remark about the solution \eqref{solution1chiral}
that we have found. From the results \eqref{Smasses} and \eqref{PSmasses}
we see that this stationary point is a minimum of the potential,
provided that $\lpq\bpq > c_1 F_X$; otherwise, the Hessian evaluated
at the stationary point would not be positive definite, being
$\frac{\de^2 \tilde{V}}{\de\sigma^2}\vert_S = 
\frac{\de^2 \tilde{V}}{\de\pi_a^2}\vert_S =
\frac{1}{2}(\lpq\bpq-c_{1}F_{X}) < 0$.
Remembering that, for $\tro < T < \tuone$,
$\rho_\pi \equiv -\frac{1}{2} \bpq < 0$, the condition for the stationary
point \eqref{solution1chiral} to be a minimum can be written as
\begin{equation}\label{Gpi}
\Gpi \equiv c_1 F_X + 2\lpq \rho_\pi = c_1 F_X - \lpq\bpq < 0 ,~~~~
{\rm i.e.:}~~~~ \rho_\pi < -\frac{c_1 F_X}{2\lpq} .
\end{equation}
In other words, assuming $c_1 F_X > 0$ and approximately constant (as a
function of the temperature $T$) around $\tro$, we have that the stationary
point \eqref{solution1chiral} is a solution, i.e., a minimum of the potential,
not immediately above $\tro$, where the parameter $\rho_\pi$ vanishes
(see Table \ref{table1}) and $\Gpi$ is positive, but (assuming that $\lpq\bpq$
becomes large enough increasing $T$, starting from $\lpq\bpq=0$ at $T=\tro$)
only for temperatures that are sufficiently higher than $\tro$, so that
the condition \eqref{Gpi} is satisfied, i.e., only for $T > \tc$, where
$\tc$ is defined by the condition $\Gpi(T=\tc) = 0$,
and it is just what we can call the {\it chiral transition temperature}.
In fact, for $T>\tc$ the solution \eqref{solution1chiral} is valid,
and the $SU(2)_L \otimes SU(2)_R$ chiral symmetry is restored.
Therefore, differently from the case $N_f\geq 3$ discussed in the previous
section, where $\tc\equiv\tro$, we have here that $\tc>\tro$.

Now the question is as follows: What happens for $\tro < T < \tc$?

\noindent
{\bf C. Study of the solution for $\tro<T<\tc$}

One immediately sees that, when
\begin{equation}\label{GpibelowTch}
\Gpi \equiv c_1 F_X + 2\lpq \rho_\pi = c_1 F_X - \lpq\bpq \geq 0 ,
\end{equation}
Eqs. \eqref{minUalpha} also admit the solution (in the chiral limit $m_u=m_d=0$)
\begin{equation}\label{solution2}
\smin = \frac{1}{\Lambda_{\pi}}\sqrt{c_{1}\sqrt2\overline{\alpha}
-\lambda_{\pi}^{2}B_{\pi}^{2}} \equiv \sigma_0,~~~~ \dmin = 0,
\end{equation} 
with $\ov{\alpha}$ defined implicitly by the third equation in Eqs.
\eqref{minUalpha}, i.e.,
\begin{equation}\label{alphasol2}
\lxq\left(\ov{\alpha}^{2}-\frac{F_X^2}{2}\right)\ov\alpha =
\frac{c_1}{2\rad2 \Lpq}(c_1\rad2\ov{\alpha}-\lpq\bpq) .
\end{equation}
This solution, being of the form $\ov{U} = \frac{\sigma_0}{\rad2}{\bf I}$, with
$\sigma_0>0$, spontaneously breaks the chiral symmetry down to the vectorial
subgroup $U(2)_V$.
It is easy to verify that, by virtue of the condition \eqref{GpibelowTch},
Eq. \eqref{alphasol2} admits a unique solution such that
$\ov\alpha \geq \frac{F_X}{\rad2} \geq \frac{\lpq\bpq}{\rad2 c_1}$
[where the last inequality comes from the condition \eqref{GpibelowTch}],
thus leading to a well-defined solution \eqref{solution2} for $\smin$.
When, in particular, $\Gpi=0$ (i.e., when $T=\tc$), then the solution
coincides with Eq. \eqref{solution1chiral}, being $\ov\alpha=\frac{F_X}{\rad2}$
and $\smin=\dmin=0$. Instead, for $\Gpi>0$ (i.e., for $T<\tc$),one has that
$\amin>\frac{F_X}{\rad2}$ and $\smin > 0$.
By studying the matrix of the second derivatives of the potential,
calculated in this stationary point, one immediately verifies that this
solution is a minimum of the potential and that the masses of
the pseudoscalar $q\bar q$ excitations $\pi_a$ (the {\it pions}) vanish; i.e.,
the $\pi_a$ are the three Goldstone bosons coming from the breaking of
$SU(2)_L \otimes SU(2)_R$ down to $SU(2)_V$.
Obviously, the solution \eqref{solution2}--\eqref{alphasol2} continues
to be valid also for $T<\tro$, where $\ropi\equiv\frac{\apq}{2}>0$,
provided that one substitutes $\bpq$ with $-2\ropi = -\apq$.

\noindent
{\bf D. Chiral condensate for $T>\tc$ and for $T<\tc$}

\begin{figure}[htbp]
\begin{center}
\includegraphics{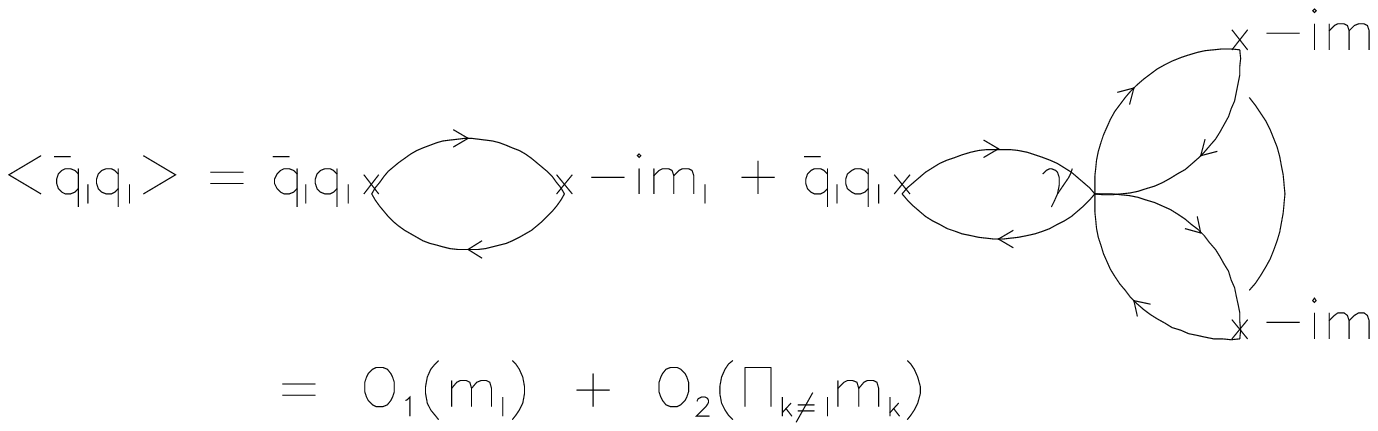}
\caption{The chiral condensate above $\tc$.}\label{fig1}
\end{center}
\end{figure}

It is well known that, since the derivative of the QCD Hamiltonian with respect
to the quark mass $m_l$ is the operator $\ov{q}_l q_l$ (being
$\delta\mathcal{L}_{QCD}^{(mass)} = -\sum_{l=1}^{N_f} m_l \ov{q}_l q_l$),
then the corresponding derivative of the vacuum energy represents the vacuum
expectation value of $\ov{q}_l q_l$, i.e., the so-called chiral
condensate. In terms of the effective Lagrangian, this means
\begin{equation}\label{chiralcondensate}
\bra\ov{q}_lq_l\ket=\frac{\de \ov{V}}{\de m_l} ,
\end{equation}
where $\ov{V}=\tilde{V}(\ov{U},\ov{U}^{\dagger},\ov{X},\ov{X}^{\dagger})
=V(\ov{U},\ov{U}^{\dagger},\ov{X},\ov{X}^{\dagger})$
is the vacuum expectation value of the potential of the effective Lagrangian.
Using the fact that $\ov\beta = \ov\eta = \ov\pi_a = \ov\delta_{1}
= \ov\delta_{2} = 0$, we find, from Eqs. \eqref{Vanom} and \eqref{VL2},
\begin{equation}\label{Vmin}
\begin{split}
\ov{V}=&~\frac{1}{8}\lpq B_{\pi}^4
+\frac{1}{8}\Lambda_{\pi}^{2}(\overline{\sigma}^{2}+\overline{\delta}^{2})^{2}
+\frac{1}{2}\lambda_{\pi}^{2}\overline{\sigma}^2\overline{\delta}^{2}
+\frac{1}{4}\lambda_{X}^{2}\left(\alpha^{2}-\frac{F_{X}^{2}}{2}\right)^{2}\\
&+\frac{1}{4}(\lambda_{\pi}^{2}B_{\pi}^{2}-\sqrt{2}c_1\ov\alpha)\ov\sigma^2
+\frac{1}{4}(\lambda_{\pi}^{2}B_{\pi}^{2}+\sqrt{2}c_1\ov\alpha)\ov\delta^2 \\
&-\frac{B_{m}}{2}[(m_{u}+m_{d})\overline{\sigma}+(m_{u}-m_{d})\overline{\delta}]
,
\end{split}
\end{equation}
which, when inserted into Eq. \eqref{chiralcondensate}, gives
\begin{eqnarray}\label{chiralcondensate-updown}
\langle\overline{q}_{u}q_{u}\rangle = \frac{\partial \ov{V}}{\partial m_{u}}
= -\frac{B_{m}}{2}(\overline{\sigma}+\overline{\delta}) ,\nonumber \\
\langle\overline{q}_{d}q_{d}\rangle = \frac{\partial \ov{V}}{\partial m_{d}}
= -\frac{B_{m}}{2}(\overline{\sigma}-\overline{\delta}) ,
\end{eqnarray}
having used Eqs. \eqref{minUalpha} for the vacuum expectation values
$\ov\sigma$, $\ov\delta$, and $\ov\alpha$.
Substituting the solutions \eqref{solution1} into the expressions
\eqref{chiralcondensate-updown}, we find that, for $T>\tc$,
\begin{eqnarray}\label{condchiL2}
\bra\ov{q}_{u}q_{u}\ket &\simeq& -\frac{B_m^2}
{\lambda_\pi^4B_\pi^4-c_1^2F_X^2}(m_u\lpq\bpq+m_dc_1F_X) , \nonumber \\
\bra\ov{q}_{d}q_{d}\ket &\simeq& -\frac{B_m^2}
{\lambda_\pi^4B_\pi^4-c_1^2F_X^2}(m_d\lpq\bpq+m_uc_1F_X) .
\end{eqnarray}
As we see, the chiral condensate vanishes in the chiral limit $m_u=m_d=0$,
signalling the restoring of the chiral symmetry. Concerning the dependence
on the quark masses, we observe that, in agreement with what was already
found in Ref. \cite{EM1994a} for $N_f\geq 3$, also for the case $N_f=2$
the expression \eqref{condchiL2} for the chiral condensate comes out to be
the sum of two contributions,
$\qq=\mathcal{O}_1(m_l)+\mathcal{O}_2(\prod_{k\neq l} m_k)$, for which the
diagrammatic interpretation is rather simple (see Fig. \ref{fig1}):
the first term $\mathcal{O}_1(m_l)$ corresponds to a diagram with the
insertion of a mass operator $-m_l \ov{q}_l q_l$, while the second term
$\mathcal{O}_2(\prod_{k\neq l} m_k)$ clearly corresponds to a diagram with the
insertion of the $2N_f$-quark effective vertex (``$\gamma$'') associated
with the $U(1)$ axial condensate $\ov{X}$.

Instead, for $T<\tc$, one finds, using the solution \eqref{solution2} (with the
substitution $\bpq \to -2\ropi \equiv -\apq$, if it is also $T<\tro$),
that, in the chiral limit $m_u=m_d=0$,
\begin{equation}
\bra\ov{q}_{u}q_{u}\ket = \bra\ov{q}_{d}q_{d}\ket
= -\frac{1}{2} B_m \sigma_0 \equiv -\frac{1}{2} B_m F_\pi,
\end{equation}
since, in this case, $\ov{U} = \frac{\sigma_0}{\rad2} {\bf I}$, and,
therefore, on the basis of what we have observed in Sec. 2 [see, in
particular, Eq. \eqref{UWDVminore} and the third footnote],
$\sigma_0$ must be identified with the {\it pion decay constant}:
$\sigma_0 \equiv F_\pi$.

\newsection{Comments on the results and conclusions}

\noindent
Let us first summarize the results that we have found.

Chiral symmetry restoration at nonzero temperature is often studied in the
framework of the effective Lagrangian \eqref{lageffstandard}--\eqref{thooftterm}
(see, e.g., Refs. \cite{PW1984,LRS2000,RRR2003,Vicari-et-al.}),
written in terms of the (quark-bilinear) mesonic effective field $U$ as (in
the chiral limit $M=0$) $\mathcal{L}_1 = \mathcal{L}_0 + \mathcal{L}_{I}$,
where $\mathcal{L}_0$ describes a kind of linear sigma model
[see Eq. \eqref{sigmamodel}], while $\mathcal{L}_{I}$ is an interaction term
of the form $\mathcal{L}_{I}=c_{I}[\det U+\det U^{\dagger}]$.
However, as was noticed by Witten \cite{Witten1980}, Di Vecchia, and Veneziano
\cite{DV1980}, this type of anomalous term
does not correctly reproduce the U(1) axial anomaly of
the fundamental theory (i.e., of the QCD), which is instead correctly
implemented in the effective Lagrangian $\mathcal{L}_2$,
written in Eq. \eqref{lageffint}, which was derived in Refs.
\cite{Witten1980,DV1980,RST1980,KO1980,NA1981}.
For studying the phase structure of the theory at finite temperature, all
the parameters appearing in the effective Lagrangian must be considered as
functions of the physical temperature $T$.
However, the anomalous term in Eq. \eqref{lageffint} makes sense
only in the low-temperature phase ($T<T_{c}$), and it is singular for
$T>T_{c}$, where the vacuum expectation value of the mesonic field $U$
vanishes. On the contrary, the interaction term $\mathcal{L}_I$
behaves well both in the low- and high-temperature phases.

To overcome the above-mentioned problems, we have considered a
modified effective Lagrangian (which was originally proposed in
Refs. \cite{EM1994a,EM1994b,EM1994c} and elaborated on in Refs.
\cite{MM2003,EM2004,EM2011}), which generalizes the two effective
Lagrangians $\mathcal{L}_1$ and $\mathcal{L}_2$ mentioned above, in such a way
that it correctly satisfies the transformation property \eqref{trasfanomal}
under the chiral group but also includes an interaction term containing the
determinant of the mesonic field $U$, of the kind of that in
Eq. \eqref{thooftterm}, assuming that there is a
$U(1)_A$-breaking condensate that (possibly) survives across the chiral
transition at $T_{c}$, staying different from zero up to a temperature
$T_{U(1)} > T_{c}$.
The modified effective Lagrangian is written in terms of
the $q\bar{q}$ mesonic effective field $U$ and of the $2N_f$-quark
(exotic) mesonic field $X$, associated with the $U(1)$ axial
condensate, and it is given by Eqs. \eqref{lagtotb}--\eqref{Vtilde}.
In particular, the potential term
$V(U,U^{\dagger},X,X^{\dagger})$, written in Eq. \eqref{V}, contains an
interaction term between the $U$ and $X$ fields, i.e.,
$\mathcal{L}_{int}=\frac{c_{1}}{2\rad2}[X^{\dagger}\det U+X\det U^{\dagger}]$,
which is very similar to the interaction term $\mathcal{L}_{I}$
that we have discussed above for the effective Lagrangian $\mathcal{L}_1$.
Even if this term is not anomalous, being invariant
under the chiral group $U(N_f)_L \otimes U(N_f)_R$, by virtue of Eqs.
\eqref{trasfU} and \eqref{trasfX},
nevertheless, if the field $X$ has a (real) nonzero vacuum expectation
value $\ov X$ (the $U(1)$ axial condensate), then, writing
$X=(\ov{X}+h_X)e^{i\frac{S_X}{\ov{X}}}$ (with $\ov{h}_X = \ov{S}_X = 0$)
and expanding in powers of the excitations $h_X$ and $S_X$,
one recovers, at the leading order, an interaction term of the form
$\mathcal{L}_{I}$: $\mathcal{L}_{int} = c_{I}[\det U+\det U^{\dagger}] + \dots$,
with $c_{I}=\frac{c_1\ov X}{2\rad2}$.
In Secs. 3 and 4 of this paper, we have analyzed in detail the effects of
assuming a nonzero value of the $U(1)$ axial condensate $\ov{X}$ on the scalar
and pseudoscalar mesonic mass spectrum above the chiral transition
temperature ($T>T_{c}$), both for the case $N_f\geq 3$ (Sec. 3) and for
the case $N_f=2$ (Sec. 4).

In particular, in the chiral limit $M=0$, one has that, for $T>T_{c}$,
$\ov U = 0$, $\ov X = \sqrt{\rho_X} \equiv \frac{F_X}{\rad2}$ [where $\rho_X
\equiv \frac{F_X^2}{2}$, see Eq. \eqref{Trange}, is the parameter
appearing in the potential term \eqref{V}], which means that the
$SU(N_f)_L\otimes SU(N_f)_R$ chiral symmetry is restored, while the $U(1)$ axial
symmetry is broken by the $U(1)$ axial condensate $\ov X$.
Concerning the mass spectrum of the effective Lagrangian, first of all we have
two exotic $2N_f$-quark mesonic excitations, described by the scalar
singlet field $h_X \sim {\det}(\bar{q}_{sL}q_{tR}) + {\det}(\bar{q}_{sR}q_{tL})$
and by the pseudoscalar singlet field
$S_X \sim i[ {\det}(\bar{q}_{sL}q_{tR}) - {\det}(\bar{q}_{sR}q_{tL})]$,
with squared masses given by $M^2_{h_X} = 2\lxq\rho_X^2 = \lxq\fxq$
and $M^2_{S_X} = \frac{A}{\ov{X}} = \frac{2A}{\fxq}$.
In particular, the physical interpretation of the pseudoscalar singlet
excitation $S_X$ is rather obvious, and it was already discussed in Ref.
\cite{EM1994a}; it is nothing but the would-be Goldstone particle
coming from the breaking of the $U(1)$ axial symmetry. In fact, neglecting 
the anomaly, it has zero mass in the chiral limit of zero quark masses.
Yet, considering the anomaly, it acquires a topological squared
mass proportional to the topological susceptibility $A$ of the pure YM theory,
as required by the Witten--Veneziano mechanism \cite{Witten1979,Veneziano1979}.

In addition, we have the usual $2N_f^2$ $q \bar{q}$ mesonic
excitations described by the field $U$.
In the case $N_f=2$, the restoration of the $SU(2)_L \otimes SU(2)_R$
chiral symmetry manifests itself in the appearance, in the mass spectrum
of the effective Lagrangian, of two $q\bar q$ chiral multiplets
$(\frac{1}{2},\frac{1}{2})$, namely, using for $U$ the parametrization
\eqref{parUX} in terms of the fields $\sigma$, $\eta$, $\vec\delta$, and
$\vec\pi$, $(\sigma,\vec\pi)$, with masses
$M_{\sigma}^{2} = M_{\pi}^{2} = \frac{1}{2}(\lpq\bpq-\sqrt{2}c_{1}\ov{X})$,
and $(\eta,\vec\delta)$, with masses
$M_{\eta}^{2} = M_{\delta}^{2} = \frac{1}{2}(\lpq\bpq+\sqrt{2}c_{1}\ov{X})$.
Instead, the squared masses of the $q\bar q$ mesonic excitations belonging
to a same $U(1)$ chiral multiplet, such as $(\sigma,\eta)$ and
$(\vec\pi,\vec\delta)$, remain split by the quantity\footnote{Since,
as we have seen in the previous section, $\sqrt{2} c_1 \ov{X} = c_1 F_X \geq 0$,
where we have also included the equality sign to take into account the limit
cases in which $c_1=0$ and/or $\ov{X}=0$ (see the discussion below),
Eq. \eqref{Msplit-bis} implies that $M_\pi \leq M_\delta$, which can be
proved to be an {\it exact} inequality in QCD (see, e.g., Ref. \cite{Smilga}
and references therein).}
\begin{equation}\label{Msplit-bis}
\Delta M_{U(1)}^2 \equiv M_{\eta}^2-M_{\sigma}^2
= M_{\delta}^{2}-M_{\pi}^{2} = \sqrt{2} c_1 \ov{X},
\end{equation}
proportional to the $U(1)$ axial condensate $\ov X = \frac{F_X}{\rad2}$.
This result is to be contrasted with the corresponding result obtained
in Sec. 3 for the $N_f\geq 3$ case, see Eq. \eqref{massesL3},
where all (scalar and pseudoscalar) $q\bar q$ mesonic excitations
(described by the field $U$) turned out to be degenerate, with squared masses
$M_U^{2}=\frac{1}{2}\lpq\bpq$.
(The result that we have obtained for $N_f\geq 3$ is in agreement with the
result that was found in Ref. \cite{BCM1996}, where simple group-theoretical
arguments were used to demonstrate that in the high-temperature
chirally restored phase of QCD with $N_f$ massless flavors, all $n$-point
correlation functions of quark bilinears with $n<N_f$ are invariant under $U(1)$
axial transformations; in particular, for $N_f\geq 3$, all two-point
correlation functions of quark bilinears are invariant under $U(1)$ axial
transformations, and, as a consequence, all $q\bar q$ mesonic excitations are
degenerate.)

This difference in the mass spectrum of the $q\bar q$ mesonic excitations
(described by the field $U$) for $T>T_{c}$ between the case $N_f=2$ and the
case $N_f\geq 3$ is due to the different role of the interaction term
$\mathcal{L}_{int} = c_{I}[\det U+\det U^{\dagger}] + \dots$ in the two cases.
When $N_f=2$, this term is (at the lowest order) quadratic in the
fields $U$ so that it contributes to the squared mass matrix.
Instead, when $N_f\geq 3$, this term is (at the lowest order) an interaction
term of order $N_f$ in the fields $U$ (e.g., a {\it cubic} interaction term
for $N_f=3$) so that, in the chiral limit, when $\ov{U}=0$, it does not
affect the masses of the $q\bar q$ mesonic excitations.

Alternatively, we can also explain the difference by using a ``diagrammatic''
approach, i.e., by considering, for example, the diagrams that contribute
to the following quantity $\mathcal{D}_{U(1)}$, defined as the difference
between the correlators for the $\delta^+$ and $\pi^+$ channels:
\begin{eqnarray}\label{DU1}
\lefteqn{
{\cal D}_{U(1)}(x) \equiv \langle T O_{\delta^+}(x) O_{\delta^+}^\dagger(0)
\rangle - \langle T O_{\pi^+}(x) O_{\pi^+}^\dagger(0) \rangle } \nonumber \\
& & = 2 \left[ \langle T \bar{u}_R d_L(x) \ \bar{d}_R u_L(0) \rangle
+ \langle T \bar{u}_L d_R(x) \ \bar{d}_L u_R(0) \rangle \right] .
\end{eqnarray}
What happens below and above $T_{c}$?
For $T<T_{c}$, in the chiral limit $m_1 = \dots m_{N_f} = 0$, the left-handed
and right-handed components of a given light quark flavor can be connected
through the $q\bar{q}$ chiral condensate, giving rise to a nonzero
contribution to the quantity ${\cal D}_{U(1)}(x)$ in Eq. \eqref{DU1}.
But for $T>T_{c}$, the $q\bar{q}$ chiral condensate is zero, and, therefore,
also the quantity ${\cal D}_{U(1)}(x)$
should be zero for $T>T_{c}$, {\it unless} there is a nonzero
$U(1)$ axial condensate $\ov{X}$; in that case, one should also consider
the diagram with the insertion of the $2N_f$-quark effective vertex
($\gamma$: see Fig. \ref{fig1} in Sec. 4) associated with the
$U(1)$ axial condensate $\ov{X}$.
For $N_f=2$, all the left-handed and right-handed
components of the {\it up} and {\it down} quark fields in Eq. \eqref{DU1}
can be connected through the four-quark effective vertex $\gamma$,
giving rise to a nonzero contribution to the quantity ${\cal D}_{U(1)}(x)$.
Instead, for $N_f=3$ the six-quark effective vertex $\gamma$ also generates
a couple of right-handed and left-handed {\it strange} quarks, which,
for $T>T_{c}$, can only be connected through the mass operator
$-m_s \ov{q}_s q_s$, so that (differently from the case $N_f=2$)
this contribution to the quantity ${\cal D}_{U(1)}(x)$
should vanish in the chiral limit; this implies that, for $N_f=3$ and $T>T_{c}$,
the $\vec\delta$ and $\vec\pi$ correlators are identical,
and, as a consequence, also $M_\delta = M_\pi$.
This argument can be easily generalized to include also the other meson
channels and to the case $N_f>3$.

Finally, let us see how our results for the mass spectrum compare with
the available lattice results.

As we have already said in the introduction,
information on the mass spectrum of the $q\bar{q}$ mesonic excitations
of the theory can be obtained by studying the two-point correlation functions
of proper quark-bilinear operators: lattice results for the case $N_f=2$
already exist in the literature, even if the situation is, at the moment, a bit
controversial. In fact, there are lattice results
\cite{lat1997,lat1998,lat1999,lat2000,lat2000bis,lat2011,lat2012,lat2013},
some of them obtained using the so-called {\it staggered fermions} on the
lattice and some others using the so-called {\it domain-wall fermions} on
the lattice, which indicate the {\it nonrestoration} of the $U(1)$ axial
symmetry above the chiral transition at $T_c$, in the form of a small (but
nonzero) splitting between the $\vec\delta$ and $\vec\pi$ correlators above
$T_c$, up to $\sim 1.2~T_{c}$.\footnote{We must point out that some of the
above-mentioned lattice results \cite{lat2011,lat2012,lat2013} refer,
properly speaking, neither to the case $N_f=2$ nor to the case $N_f=3$ but
to the (more realistic) case ``$N_f=2+1$,'' in which there are two ({\it up}
and {\it down}) very light (eventually massless) quark flavors and one
massive {\it strange} quark with a realistic mass $m_s \sim 100$ MeV.
However, it is commonly believed (see, e.g., Refs. \cite{Vicari-et-al.} and
references therein) that, due to the large mass of the {\it strange} quark,
this case, at least in the vicinity of the phase transition at $T_c$, is closer
to the ideal case $N_f=2$ (obtained in the limit $m_s \to \infty$) rather than
to the ideal case $N_f=3$ (obtained in the limit $m_s \to 0$). Moreover, the
fact that also, in this case, a splitting is observed between the $\vec\delta$
and $\vec\pi$ correlators immediately above $T_c$ can be considered (on the
basis of our previous arguments) as an {\it a posteriori} confirmation of this
expectation.}
In terms of our result \eqref{Msplit-bis}, we would interpret this by saying
that, for $T>\tc$, there is still a nonzero $U(1)$ axial condensate,
$\ov{X} > 0$, so that $c_{I}=\frac{c_1\ov X}{2\rad2} > 0$ and the
above-mentioned interaction term, containing the determinant of the mesonic
field $U$, is still effective for $T>\tc$.

However, recently, other lattice results, obtained using the so-called
{\it overlap fermions} on the lattice, have been reported \cite{Cossu2013},
which do not show evidence of the above-mentioned splitting above $\tc$, so
indicating an {\it effective} restoration of the $U(1)$ axial symmetry
above $\tc$, at least, at the level of the $q\bar{q}$ mesonic mass spectrum
(see also Ref. \cite{Aoki2012}, where the same conclusions have been derived
analytically but always using the overlap lattice fermions, with the
help of certain assumptions).
In terms of our result \eqref{Msplit-bis}, we would interpret this by saying
that, for $T>\tc$, one has $c_1 \ov{X} = 0$, so that
$c_{I}=\frac{c_1\ov X}{2\rad2} = 0$ and the above-mentioned interaction term,
containing the determinant of the mesonic field $U$, is not present
for $T>\tc$.
For example, it could be that also the $U(1)$ axial condensate $\ov{X}$
(like the usual chiral condensate $\langle \bar{q} q \rangle$) vanishes
at $T=\tc$, i.e., using the notation introduced in Sec. 2
(see Table \ref{table1}), that $\tuone=\tc$. (Or, even more drastically,
it could be that, at least for $N_f=2$, there is simply {\it no} genuine
$U(1)$ axial condensate \dots .)
In this case, in order to preserve the consistency of our
effective model, we should require that also the pure-gauge topological
susceptibility $A(T)$ vanishes immediately above the critical temperature $\tc$;
otherwise, the anomalous term in Eq. \eqref{Vtilde} would be singular above the
critical temperature $T_{c}$, where the vacuum expectation values of the
mesonic fields vanish (in the chiral limit $M=0$).
However, lattice results show that the pure-gauge topological susceptibility
$A(T)$ is approximately constant up to the critical temperature $T_{c}$, and
then it has a sharp decrease above the transition, but it remains different
from zero, at least up to $\sim 1.2~T_{c}$ (this suppression for $T>\tc$,
however, increases when increasing the number $N_c$ of colors, thus hinting
at a vanishing large-$N_c$ limit of $A(T)$ for $T>\tc$, as it was suggested
in Ref. \cite{KPT1998}. See Ref. \cite{VP-report} for a recent review on
these problems.)
We recall that, in the Witten--Veneziano mechanism
\cite{Witten1979,Veneziano1979}, a (no matter how small) value
different from zero for $A$ is related to the breaking of the $U(1)$ axial
symmetry, since it implies the existence of a pseudoscalar and
flavor-singlet would-be Goldstone particle; thus, a (small)
nonzero value of $A(T)$ for $T>\tc$ should imply a (presumably small)
nonzero value of the $U(1)$ axial condensate $\ov{X}$.

Alternatively, one could of course explain the (possible) vanishing of
the coefficient $c_{I}=\frac{c_1\ov X}{2\rad2}$ of the interaction term
containing the determinant of the mesonic field $U$ above $\tc$ simply by
assuming that the coefficient $c_1$ (possibly) vanishes above $\tc$.
(The possibility that $c_1 \equiv 0$
at every temperature $T$, including $T=0$, must be discarded if we also
assume that there is a genuine nonzero $U(1)$ axial condensate $\ov{X}$,
since, as it was shown in Appendix B of Ref. \cite{EM2011},
this hypothesis would lead to wrong predictions for the pseudoscalar-meson
mass spectrum at $T=0$.)

In conclusion, further work will be necessary, both from the analytical point
of view but especially from the numerical point of view (i.e., by lattice
calculations), in order to unveil the persistent mystery of the fate of the
$U(1)$ axial symmetry at finite temperature.

\section*{\large\bf Acknowledgments}
A. Mord\`a has been supported by the OCEVU Labex (Grant No. ANR-11-LABX-0060)
and by the A*MIDEX project (Project No. ANR-11-IDEX-0001-02), funded by the
``{\it Investissements d'Avenir}'' French government program managed by the ANR.

\newpage

\renewcommand{\Large}{\large}

\end{document}